\documentclass[twocolumn,pra,aps,superscriptaddress]{revtex4-1}

\usepackage{epsfig}
\usepackage{color}
\usepackage{physics}
\usepackage{amsmath}
\usepackage{amsfonts}
\usepackage{hyperref}

\begin{document}

\title{Hilbert-Schmidt distance and entanglement witnessing}
\author{Palash Pandya}\affiliation{Institute of Theoretical Physics and Astrophysics, Faculty of Mathematics, Physics, and Informatics,\\ University of Gda\'nsk, 80-308 Gda\'nsk, Poland}
\author{Omer Sakarya}\affiliation{Institute of Informatics, Faculty of Mathematics, Physics, and Informatics,\\ University of Gda\'nsk, 80-308 Gda\'nsk, Poland}
\author{Marcin Wie\'sniak}\affiliation{Institute of Theoretical Physics and Astrophysics, Faculty of Mathematics, Physics, and Informatics,\\ University of Gda\'nsk, 80-308 Gda\'nsk, Poland}\affiliation{International Centre for Theory of Quantum Technologies,\\ University of Gda\'nsk, 80-308 Gda\'nsk, Poland}
\begin{abstract}
Gilbert proposed an algorithm for bounding the distance between a given point and a convex set. In this article we apply the Gilbert's algorithm to get an upper bound on the Hilbert-Schmidt distance  between a given state and the set of separable states. While Hilbert Schmidt Distance does not form a proper entanglement measure, it can nevertheless be useful for witnessing entanglement. We provide here a few methods based on the Gilbert's algorithm that can reliably qualify a given state as strongly entangled or practically separable, while being computationally efficient. The method also outputs successively improved approximations to the Closest Separable State for the given state. We demonstrate the efficacy of the method with examples.
\end{abstract}
\maketitle
Entanglement is by far, the most surprising feature of quantum mechanics. For pure states, this means that the state cannot be written as tensor product of pure states as subsystems. This implies that the properties of one subsystem are defined only in reference to others, and can be established by measurements constructed on remote parts of the system. Mixed entangled states are those, which cannot be written as a mixture of pure product states. This phenomenon is the ground for advantages of various quantum information processing protocols, such as quantum communication complexity reduction protocols, or quantum games, over their classical counterparts. This advantage is often related to the Bell theorem \cite{BELL}, in which strictly quantum nature is contrasted not with local quantum-mechanical statistics, but with local realistic theories, in which outcomes of all local measurements are preassigned.

Entanglement can be classified in many different ways. The most obvious criterion is the number of subsystems effectively involved, i.e., how many subsystems need to be entangled to create the state of our interest. This classification can be generalized in at least two different ways, i.e., depth of entanglement, which tells us how many subsystems must be entangled at least to recreate the state, and the structure of entanglement, describing the necessary connections between individual constituents. For multipartite case, we have individual classes of entangled states, such that we cannot transform between these classes with local operations and classical communication (LOCC) \cite{ENTCLASSES}. We can also distinguish those entangled states, which can provide statistics violating any Bell inequalities, or those, from which maximally entangled states can be distilled by means of local operations, i.e., free entangled states, as opposed to those with bound entanglement\cite{BOUND}. Bound entangled states can exist for any system larger than $2\times 3$.

This also brings us to the problem of entanglement measures \cite{MEAS}. It is natural to quantize non-classicality. Any proper measure of entanglement must satisfy certain axioms, such as nullification for separable states, normalization for the two-qubit maximally entangled state, additivity under tensor product, or monotonicity under local operations supported by classical communication \cite{MEAS1}. For pure states, there is a unique measure given by the entropy of squares of moduli of the Schmidt coefficients. For generic states, two measures were operationally induced, i.e., distillable entanglement and entanglement cost \cite{DIST, COST}. Subsequently, other measures were proposed, but all of them suffer from the practical impossibility of calculating the value for a given state.

In such a case, a more relaxed approach is taken. Namely, it often suffices to certify that the state is entangled. This can be conveniently done with a witness operator \cite{WITNESS,DOHERTY}, which assumes mean values from a certain range, but has eigenvalues beyond this range. By Jamio\l kowski-Choi isomorphism \cite{CHOI}, they are related to positive, but not completely positive maps. This is a universal method to certify entanglement for any non-separable state. A special case of such a map is the positive partial transposition (PPT) criterion \cite{PPT}, which was subsequently generalized to symmetric extensions \cite{DOHERTY}, leading to a general entanglement criteria. Unfortunately, we do not know a general form of a witness, or all possible not completely positive maps. Likewise, quadratic entanglement criteria have been proposed, but it is also very difficult to construct a non-linear criterion for generic states.

Here, we discuss the Gilbert algorithm for estimating and bounding from above, the Cartesian distance between a given point and a convex set with a known boundary. This is known as the Hilbert-Schmidt distance (HSD) between a given state and the set of separable states. 
This is done by an infinite series of corrections. Up to now, it has been successfully used to find examples falsifying local realistic models \cite{Vertesi}, as well as to truncate quantum states so that they remain in classes of SLOCC-equivalence \cite {Guehne}. We propose a few methods to efficiently qualify a given state as strongly entangled or practically separable by analyzing the numerical output of the procedure. To apply the Gilbert algorithm, it is only necessary to be able to reach all extremal points of the set, which is easily doable for local realistic models and separable states. 

The (simplified) Gilbert algorithm is as follows:

\begin{itemize}
\item{Parameters: dimensions of subsystems $d_1, d_2, ...$,}
\item{Input data: the state to be tested $\rho_0$, and any separable state $\rho_1$}
\item{Output data: the closest state found $\rho_1$, list of values of $D^2(\rho_0,\rho_1)=\Tr(\rho_0-\rho_1)^2$.} 
\end{itemize}
\begin{enumerate}
\item{Increase the counter of trials $c_t$ by 1. Draw a random pure product state $\rho_2$, hereafter called a trial state.}
\item{Run a preselection for the trial state by checking a value of a linear functional. If it fails, go back to point~1.}
\item{In case of successful preselection symmetrize $\rho_1$ with respect to all symmetries  by $\rho_0$, which respect the separability.}
\item{Find the minimum of $\Tr(\rho_0-p \rho_1-(1-p)\rho_2)^2$ with respect to $p$.}
\item{If the minimum occurs for $0\leq p\leq 1$, update $\rho_1\leftarrow p\rho_1+(1-p)\rho_2$, add the new value of $D^2(\rho_0,\rho_1)$ to the list and increase the success $c_s$ counter value by 1. }
\item{Go to step 1 until a chosen criterion HALT is met.}
\end{enumerate}

\begin{figure}
	\centering
		\includegraphics[width=7.5cm]{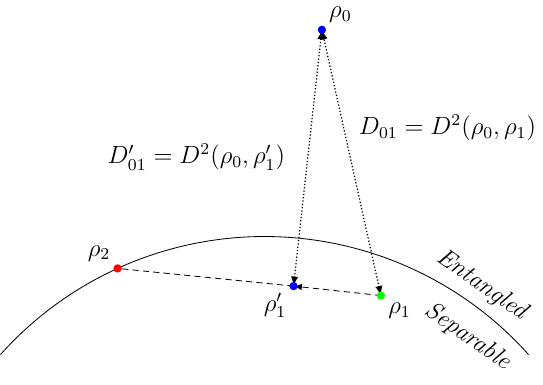}
	\caption{(color online) Visualization of the Gilbert algorithm. A random separable state $\rho_2$ is generated that satisfies the preselection criterion. Next step is to find $\rho_1' = p\rho_1+(1-p)\rho_2$, $0\leq p\leq 1$, such that the new Hilbert-Schmidt distance $D_{01}'=D^2(\rho_0,\rho_1')$ is less than the previous, $D_{01}=D^2(\rho_0,\rho_1)$. If such a state is found, the state $\rho_1'$ is updated as the new $\rho_1$. }
	\label{fig:ALG}
\end{figure}

We will now provide remarks on the algorithm.   	 

Our simplification with respect to the original version is that, while we take a random separable state, in the original version, the boundary point is optimized in each execution of the loop, i.e., we would vary $\rho_2$ to maximize $\Tr[(\rho_2-\rho_1)(\rho_0-\rho_1)]$. An example of such an implementation of the Gilbert algorithm, is in Ref \cite{Vertesi}, where the authors use an Oracle to obtain the optimized $\rho_2$ in each step. Such a maximization is by itself a semi-definite problem, and our simplified algorithm is also able to yield precise estimations. Many decompositions into product states requires many admixtures, which requires either a large number of corrections, or symmetrizations. Additionally, this simplification also contributes to the versatility of the algorithm applications.

{\em Ad. 1:} 
Instead of getting the actual Hilbert-Schmidt distance to the separable state, we get an upper bound by getting closer and closer members of this set in each step. As a result, the Gilbert algorithm cannot be used directly to certify entanglement. Nevertheless, this information can be still useful in a few different ways. First, knowing Closest Separable State (CSS), it is straight-forward to construct an entanglement witness, which reads $\rho_0 -\rho_{CSS}$. It may happen that a state found by the algorithm is close enough that it can be used to construct a successful entanglement criterion. In some cases, if both $\rho_0$ and CSS have analytical forms, it can be possible to anticipate the latter. Such an anticipation shall also be tested with the algorithm both as $\rho_0$ and initial $\rho_1$ to confirm its separability and proximity to the tested state. Finally, we can estimate $D^2(\rho_0)$ from the decay of $D^2(\rho_0,\rho_1)$. The examples calculated for this work show that there is a linear dependence between $c_s$ and $D^2(\rho_0,\rho_1)$. Namely, the sample correlation coefficient,
\begin{equation}
R(x,y)=\frac{\ev{x y}-\ev{x}\ev{y}}{\sqrt{\left(\ev{x^2}-\ev{x}^2\right)\left(\ev{y^2}-\ev{y}^2\right)}}
\end{equation}
between $x$ being the series of values of $c_s$, i.e. a sequence of consecutive positive integers, and
\begin{equation}
y_{c_s} = \abs{ \ln[D_{c_s}^2(\rho_0,\rho_1)-a]}^b,
\end{equation}
where $a$ and $b$ are free parameters of maximization of $R(x,y)$. $a\approx\lim_{c_s\rightarrow\infty}D^2(\rho_0,\rho_1)$ is the estimate for $D^2(\rho_0)$. In different cases the found values of $b$ vary, even for individual runs of the algorithm for the same state. Still, for sufficiently large values of $c_s$, the dependence is remarkably close to linear. At the same time, there is a strong linear  dependence between $c_s$ and $c_t^f$. 

Finally, it often seems from the algorithm that the CSS commutes with $\rho_0$. If this conjecture is taken, then one may observe the convergence of the eigenvalues of $\rho_1$ to some limit. Since the Hilbert Schmidt norm is invariant under any unitary, for any matrix $A$,
\begin{align}
&\Tr[(UAU^\dagger)(UAU^\dagger)^\dagger]\nonumber\\
=&\Tr[UAU^\dagger UA^\dagger U^\dagger]\nonumber\\
=&\Tr[UAA^\dagger U^\dagger]\nonumber\\
=&\Tr[AA^\dagger],
\end{align} 
we can easily find $D^2(\rho_0)$ with the limit of the spectrum of $\rho_1$. 

Since we cannot have presumptions about CSS, one shall draw states, the product of which will constitute a trial state, with the Hilbert-Schmidt distance uniform measure \cite{Karol}. To get a $d$-dimensional state, one takes a list of $2d$ random real numbers drawn with the Normal distribution with a fixed deviation and the mean equal to 0. Consecutive pairs in the list are combined to form a list of $d$ complex numbers. Subsequently, we normalize the list. If the random real numbers are evenly distributed on interval $[0,1]$, we take two of them, $x_1$ and $x_2$, and we build a normally distributed complex variable $e^{2\pi i x_1}\sqrt{-2\ln x_2}$.

When $\rho_0$ is strictly real, one may be tempted to draw real trial states. However, if one does so, misleading results are obtained. For example, for a two-qubit maximally entangled state, we get
\begin{equation}
\lim_{c_s\rightarrow\infty}\rho_{1,real}=\frac{1}{8}\left(\begin{array}{cccc}3&0&0&1\\0&1&1&0\\0&1&1&0\\1&0&0&3\end{array}\right),
\end{equation} 
while if we allow complex trial states, we get the Werner state,
\begin{equation}
\lim_{c_s\rightarrow\infty}\rho_{1,complex}=\frac{1}{6}\left(\begin{array}{cccc}2&0&0&1\\0&1&0&0\\0&0&1&0\\1&0&0&2\end{array}\right).
\end{equation}

However, there are certain states, for which taking strictly real trial states seems to suffice, which greatly reduces the complexity of the problem. 

{\em Ad 2:} The preselection criterion is $\Tr[(\rho_2-\rho_1)(\rho_0-\rho_1)]>0$. The geometrical interpretation is that the angle between vectors $\rho_0-\rho_1$ and $\rho_2-\rho_1$ is not larger than $\frac{\pi}{2}$. This implies that the point belonging to line $\{\rho_1,\rho_2\}$ lies towards $\rho_2$, i.e. $\left.\partial_p D^2(\rho_0,p\rho_1+(1-p)\rho_2)\right|_{p=0}>0$. Thus, an admixture of $\rho_2$ is certain to decrease $D^2(\rho_0,\rho_1)$. In comparison to the original Gilbert's algorithm, more corrections are made, but each of them may be less efficient.


{\em Ad 3:} If a problem possesses a symmetry, either there is a group of solutions together respecting this symmetry, or there is a unique solution possessing this symmetry. In case of the Hilbert-Schmidt distance between any given state and a convex set of separable states, we always get a unique solution.  For example, consider two states $\rho$ and $\rho'$, which are equidistant to $\rho_0$, $D^2(\rho_0,\rho)=D^2(\rho_0,\rho')$. Now, consider an arbitrary combination $((1+x)\rho+(1-x)\rho')/2$, and we get that the second derivative with respect to $x^2$ is $2D^2(\rho,\rho')>0$, while, since in our case $\rho$ and $\rho'$ are related to each other by a unitary transformation related to the symmetry of $\rho_0$, the first derivative vanishes. Hence the state minimizing $D^2$ is $(\rho+\rho')/2$. A similar argument holds for any number of equidistant states. Since there is a unique CSS, it must posses symmetries of $\rho_0$ respecting separability, we also expect $\rho_1$ to have them. Then, with $U$ representing a transformation associated with such a symmetry we have
\begin{align}
&\Tr[(U\rho_2  U^\dagger-\rho_1)(\rho_0-\rho_1)]\nonumber\\
=&\Tr[ U(\rho_2-\rho_1)U^\dagger(\rho_0-\rho_1)]\nonumber\\
=&\Tr[ U(\rho_2-\rho_1)U^\dagger U(\rho_0-\rho_1)U^\dagger]\nonumber\\
=&\Tr[ (\rho_2-\rho_1)(\rho_0-\rho_1)].
\end{align}
In the algorithm, if the state $\rho_0$ possesses a discrete symmetry, $U$ of order $k$, then the preselected $\rho_2$ is symmetrized according to $\rho_2\leftarrow \frac{1}{k}\sum_{z=0}^{k-1}[U^z\rho_2(U^\dagger)^z]$, and likewise in the case of continuous symmetry, the summation is replaced by an integral.


Let us now present few examples of our algorithm. First, we consider the maximally entangled state of two $d$-dimensional system,
\begin{equation}
\ket{\phi_d}=\frac{1}{\sqrt{d}}\sum_{i=0}^d\ket{i,i}.
\end{equation}
This a the special case, as we know the CSS explicitly, the $d$-dimensional Werner state. It is the mixture of $\op{\phi_d}$ and the white noise with respective weights $\frac{1}{d+1}$ and $\frac{d}{d+1}$. This give $D^2(\op{\phi_d})=\frac{d-1}{d+1}$. In Fig.~2 we present the convergence of $D^2(\rho_0,\rho_1)$ to $\frac{d-1}{d+1}$ for $d=2,3,\dots,9$.
\begin{figure}[tbh!]
	\centering
		\includegraphics[width=8cm]{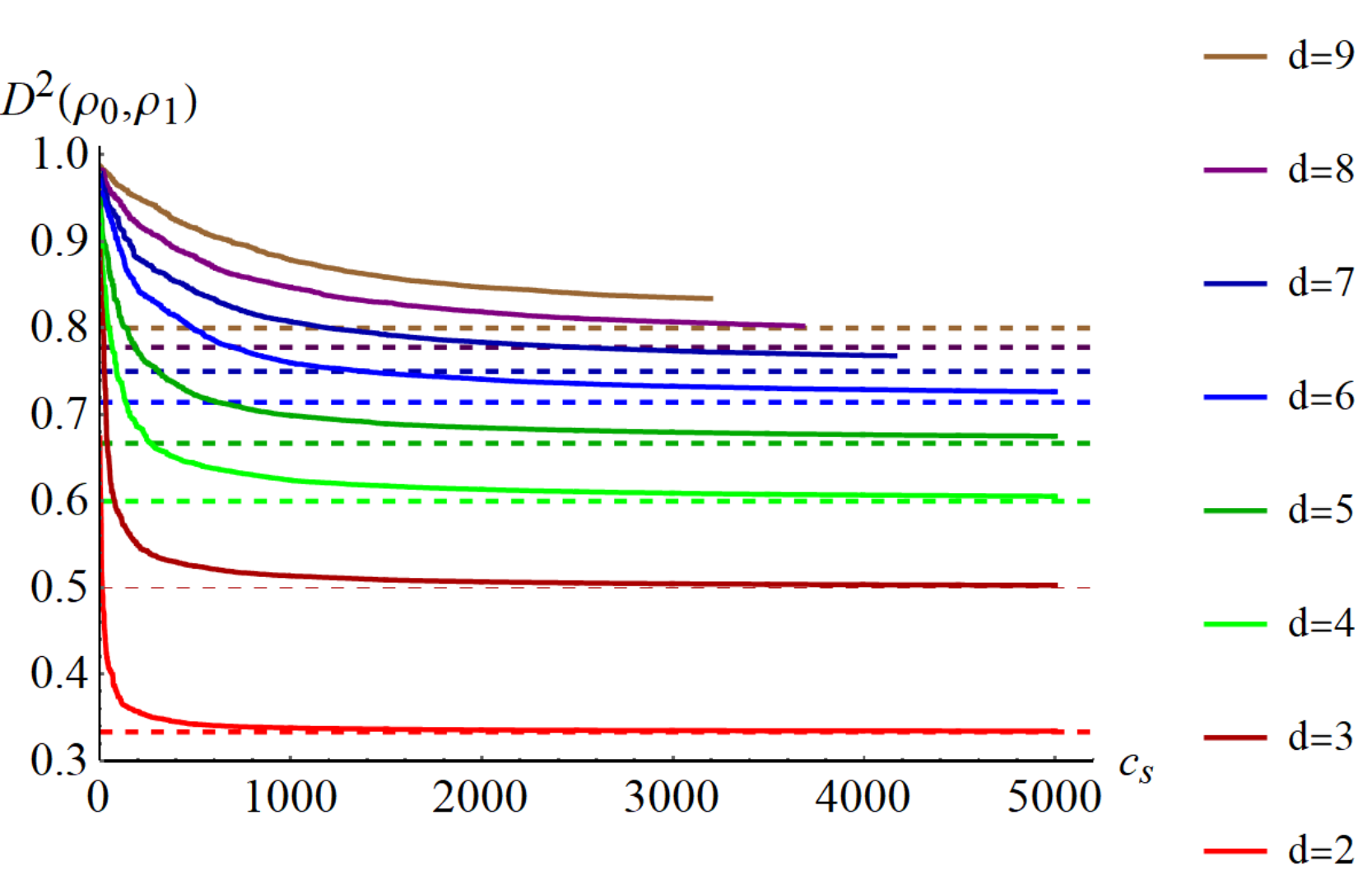}
	\caption{(color online) Convergence of $D^2(\rho_0,\rho_1)$ (y-axis) to the analytical limit (dashed lines) as a function of number of corrections $c_s$ (x-axis) for $d=2,3,\dots, 9$ (bottom to top).}
	\label{fig:MES}
\end{figure}

We made 21 runs of the algorithm for the two-ququart maximally entangled state ($d=4$) with HALT condition $c_s=1000$. Series of values of $D^2(\rho_0,\rho_1)$ from each run were used to maximize the sample correlation coefficient, as described above. Each time the maximization yielded a value above 0.999. The dependence of exponent $b$ on the limit of HSD $a$ is shown in Fig.~3.

\begin{figure}[t!]
	\centering
		\includegraphics[width=7cm]{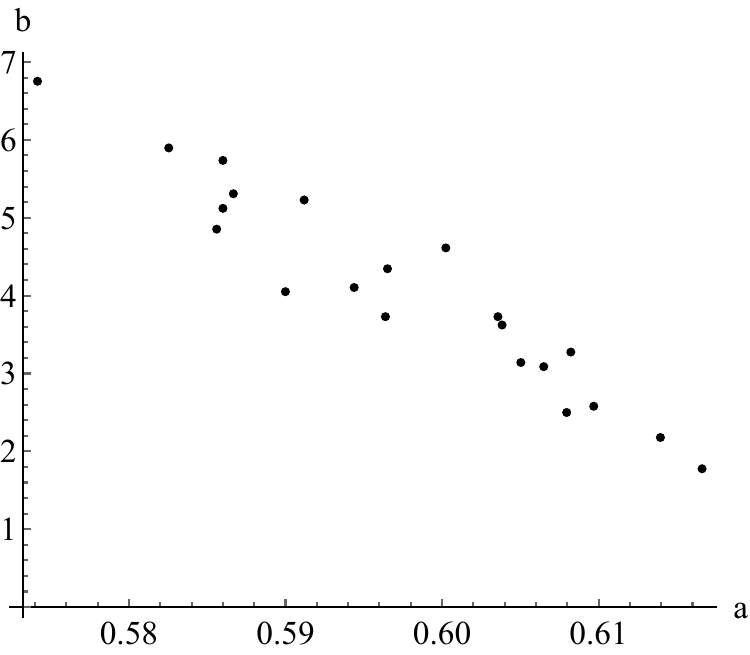}
	\caption{Relation between $a$ and $b$ found in maximizing $R(x,y)$ in 21 runs of the algorithm for the maximally entangled state of two ququarts ($d=4$) and HALT condition $c_s=1000$. In general, underestimation of the HSD is related to slow initial convergence (high values of $b$), while overestimation is accompanied by fast convergence.}
	\label{fig:SINGLET4}
\end{figure}

Another interesting case are GHZ states of $N$ qubits,
\begin{equation}
\ket{GHZ_{N,2}}=\frac{1}{\sqrt{2}}\left(\ket{0}^N+\ket{1}^N\right).
\end{equation}

\begin{figure}
 \centering
 	\includegraphics[width=8.6cm]{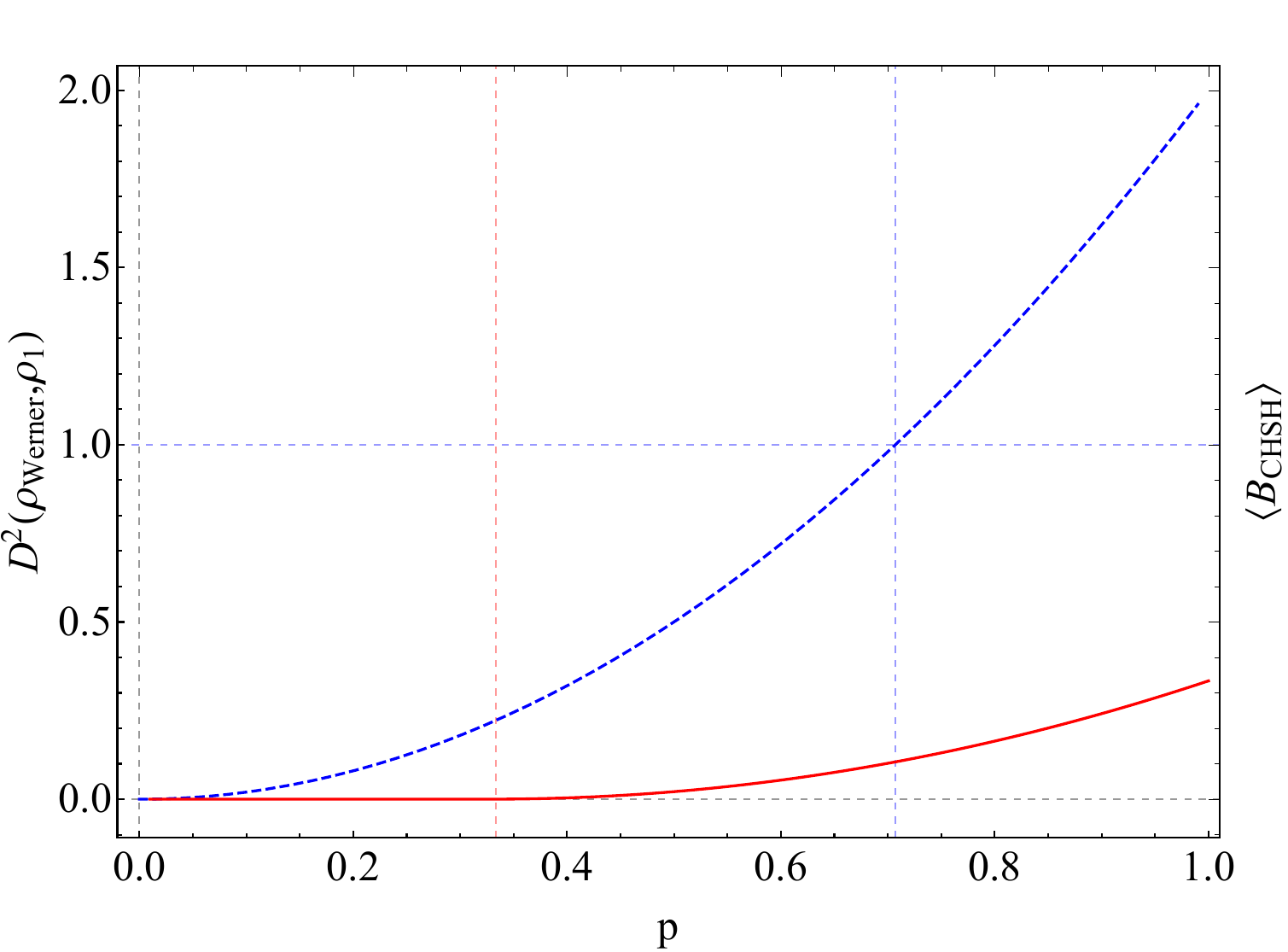}
 	\caption{(color online) Plot of the minimum distance $D^2(\rho_{\mbox{\tiny{Werner}}},\rho_1)$ (in red, solid line) for $d=2$ and expectation of CHSH operator (with 1 being the maximal local realistic values, in blue, dashed line). The CHSH inequality is violated after the minimum distance becomes non-zero.}
 	\label{fig:werner-d2}
 \end{figure}

\begin{figure}
 \centering
 	\includegraphics[width=8.6cm]{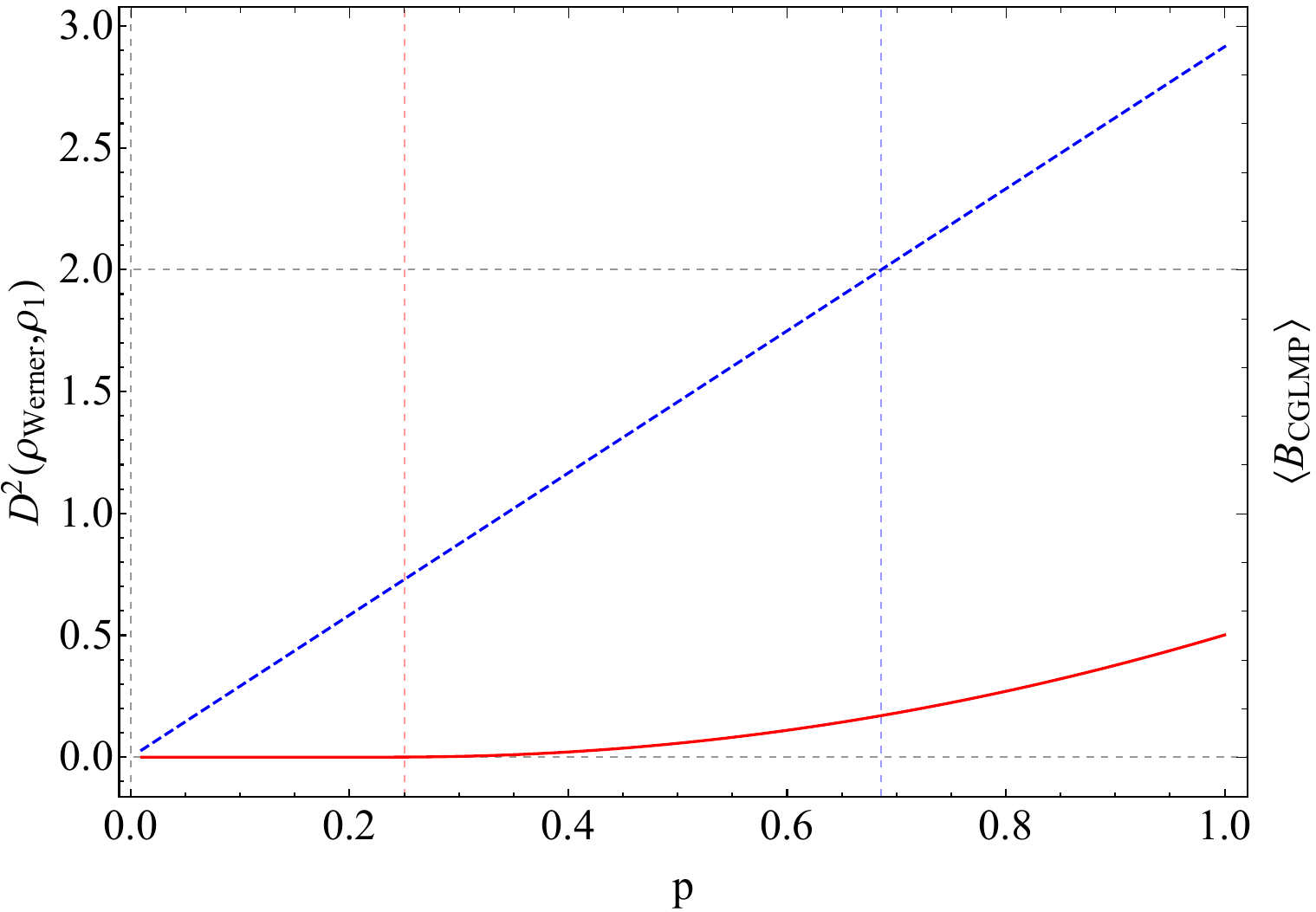}
 	\caption{(color online) Plot of minimum $D^2(\rho_{\mbox{\tiny{Werner}}},\rho_1)$ for the $d=3$ Werner state (in red, solid line) and expectation of CGLMP operator (with 2 being the maximal local realistic values, in blue, dashed) with the visibility parameter $p$.}
 	\label{fig:werner-d3}
 \end{figure}

\begin{figure}
 \centering
 	\includegraphics[width=8.6cm]{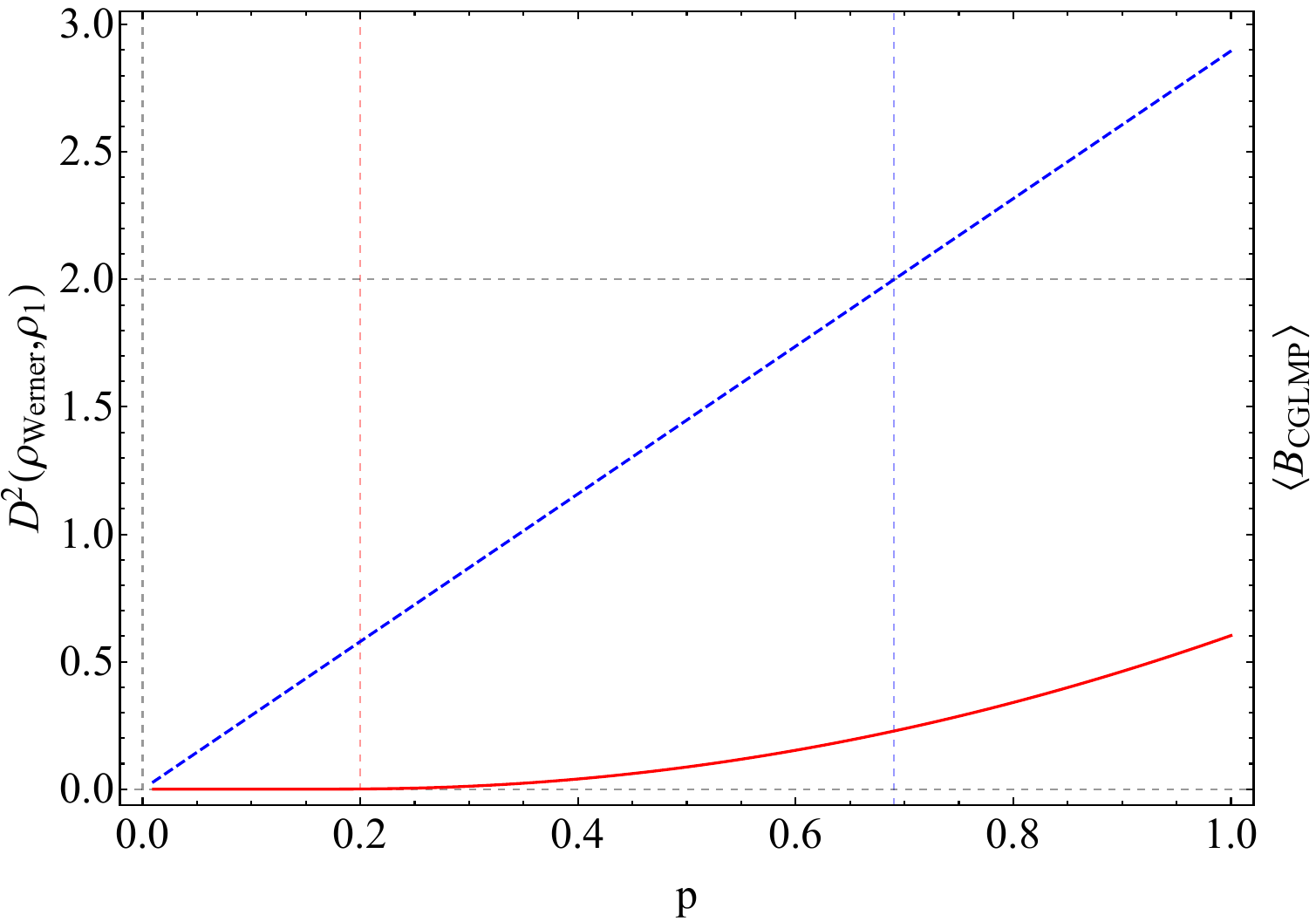}
 	\caption{(color online) Plot of minimum $D^2(\rho_{\mbox{\tiny{Werner}}},\rho_1)$ for the $d=4$ Werner state (in red, solid line) and expectation of CGLMP operator (with 2 being the maximal local realistic values, in blue, dashed) with the visibility parameter $p$.}
 	\label{fig:werner-d4}
 \end{figure}

In this case we also know the CSS, which is a mixture, $\sigma_N=x_N\sigma_{1,N}+(1-x_N)\sigma_{2,N}$, where 
\begin{align}
\sigma_{1,N}=\frac{1}{2}&\left(\begin{array}{ccccc}1&0&\ldots&0&0\\0&0&\ldots&0&0\\\vdots&\vdots&\ddots&\vdots&\vdots\\0&0&\ldots&0&0\\0&0&\ldots&0&1\end{array}\right),\nonumber\\
\sigma_{2,N}=\frac{1}{2^N}&\left(\begin{array}{ccccc}1&0&\ldots&0&1\\0&1&\ldots&0&0\\\vdots&\vdots&\ddots&\vdots&\vdots\\0&0&\ldots&1&0\\1&0&\ldots&0&1\end{array}\right),\nonumber\\
x_N=&\frac{(2^N-2)^2}{4+4^N-2^{N+1}},\nonumber\\	
D^2(\op{GHZ_{n,2}})=&\frac{(2^N-2)}{-4+2^{3-N}+2^{N+1}}.
\end{align}
 Our algorithm converges to these states, the convergence is much faster if we take a decohered GHZ state as the initial $\rho_1$. For $N=4$ and with initial $\rho_1=\sigma_N$, the algorithm was unable to find a single corrections within 6,000,000 trial states. We also considered the W states for $N=3,4$ and $5$ qubits. In each of the three cases the algorithm provides an entanglement witness with $D^2(\rho_0,\rho_1)=0.432162, 0.473063, 0.501705, 0.544239$ respectively for $N=3,4,5,6$ and number of corrections $c_s=10000, 9700, 2600, 900$. Another experiment to demonstrate the capability of entanglement detection by the algorithm is to consider Werner states for dimensions $d=2,3,4$. We compared the minimum HSD of the Werner states with the expectation of the Bell-CHSH operator for $d=2$, and minimum HSD with the expectation of the Collins-Gisin-Linden-Massar-Popescu (CGLMP) operator \cite{CGLMP} for $d=3,4$ cases for different values of the visibility parameter.
\begin{align}
\rho_{\mbox{\tiny{Werner}}} = p \op{\phi_d} +(1-p)\frac{\mathbb{I}}{d}\,,
\end{align}
where $0\le p\le 1$ is the visibility parameter and $\ket{\phi_d}$ is the maximally entangled state of $2$ qu$d$its.
In case of $d=2$, the minimum $D^2(\rho_{\mbox{\tiny{Werner}}},\rho_1)$ compared with the expectation of the CHSH operator, for a hundred values of the visibility parameter, is shown in the Figure \ref{fig:werner-d2}. It is clear that the detection of entanglement (minimum distance becomes non-zero), happens well before the CHSH violation, at $p\approx0.33$, according to the PPT criterion. We see similar results in the case of $d=3$ and $d=4$, where $D^2(\rho_{\mbox{\tiny{Werner}}},\rho_1)$ is compared to the expectation of the CGLMP operator, presented in Figures \ref{fig:werner-d3} and \ref{fig:werner-d4}. In both cases, we again see the detection of entanglement happen well before the violation of CGLMP inequality, at $p\approx\frac{1}{d+1}$, for both $d=3$ and $4$. This complements the finding that the CSS of maximally entangled state of 2 $d$-dimensional systems is the Werner state that has the visibility $\frac{1}{d+1}$.


In case of the 4-qubit cluster state, after $c_s=5000$, the minimum distance achieved was $0.632264$ compared to the predicted minimum distance of $0.491228$ for a 4 qubit GHZ state. Another case study comprised of running the algorithm for 10,000 2-qubit and 10,000 3-qubit randomly generated states. With the HALT condition set as $c_s=10000$ or $D^2(\rho_{\mbox{\tiny{Werner}}},\rho_1)\le 10^{-6}$, in case of all but 4 states, the algorithm provided a valid entanglement witness.

The last example is the two-qutrit UPB bound entangled (UPB BE) state  \cite{UPB}. With the Gilbert algorithm, we have reached $D^2(\rho_0,\rho_1)=0.002177$ with $c_s=25800$ and $c_t= 50,000,000$. We then used the the sequence of values of $D^2(\rho_0,\rho_1)$ to maximize the sample correlation coefficient, getting $R(x,y)=0.999986$ with list of values of $D^2(\rho_0,\rho_1)$ for $c_s\pmod{100}=0$, $a=0.00195735$, and $b=8.25596$. We also found a fit $c_s\propto c_t^{2.2662}$. We then ran the algorithm again with the state found previously as $\rho_1$ and the HALT condition $c_t=50 000 000$. This allowed for $c_s=64317$,  $D^2(\rho_0,\rho_1)\leq 1.382\times 10^{-5}$, $a=6.5\times 10^{-7}$ and $f=0.614$. Thus, Gilbert's algorithm can distinguish between entangled and non-entangled states with very high precision. The pace of convergence is also a sign for separability or entanglement.

We also used the results for two-qutrit UPB BE states to construct an entanglement witness. If CSS is known to be $\rho_{CSS}$, a necessary entanglement condition for state $\rho$ is $\Tr \rho(\rho_0-\rho_{CSS})>0$. However, we typically do not know $\rho_{CSS}$, but only $\rho_1$. The optimal entanglement witness for $\rho_0$ shall take form $W_{\rho_0}=(\rho_0-\rho_1)-\max_{|\phi\rangle \in \text{SEP}}\langle\phi|\rho_0-\rho_1|\phi\rangle$. After taking 20500 corrections to $\rho_1$ we get $\max_{|\phi\rangle \in \text{SEP}}\langle\phi|\rho_0-\rho_1|\phi\rangle=0.0130011$ and $\Tr\rho_0(\rho_0-\rho_1)=0.0148792$, hence leading to a valid bound entanglement witness. Ref. \cite{ThisIsUs} demonstrates that for 146 UPB BE states in dimensions $3\times 3$, $4\times 4$, $5\times 5$ and $6\times 6$, we found 123 entanglement witnesses, which are stronger than those known previously for these class \cite{BGR}.

\begin{figure}
	\centering
		\includegraphics[width=8.5cm]{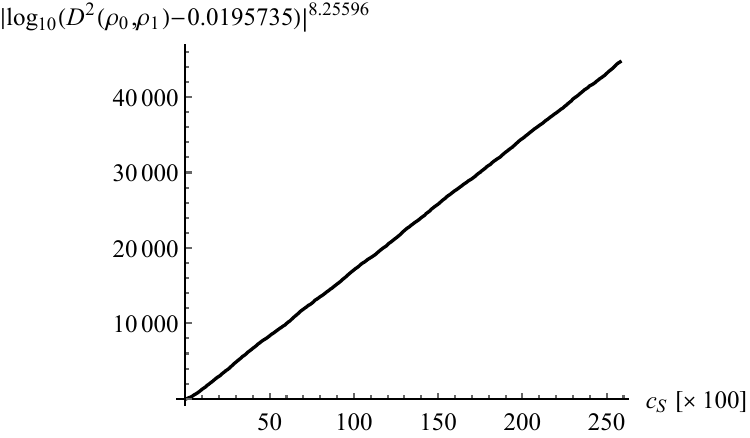}
	\caption{Dependence between the number of corrections to $\rho_1$ and $D^2(\rho_0,\rho_1)$ for the two-qutrit UPB BE state \cite{UPB}.}
	\label{fig:UPBBE}
\end{figure}

{\em Conclusions:} We have presented the use of the Gilbert algorithm to estimate the Hilbert-Schmidt distance between a given state and the set of separable states. It works by a successive construction of an optimal separable states. While the optimal state cannot be reached by the algorithm, it still gives us a lot of useful information on entanglement of a given state. The algorithm guaranties that the state will converge to a close approximation of the CSS as the number of corrections increases. It can be straight-forwardly implemented on most computational platforms, does not require large amounts of memory, and is readily formulated for any system.

It is interesting to make a brief comparison between the PPT symmetrical extension (PPTSE) method presented in Refs. \cite{DOHERTY} and the algorithm discussed here. In theory, both methods would give a definite answer about the separability of a state, they can also (with high probability) yield an entanglement witness. Let us stress a few differences. The symmetrical extension estimates the bound of set of separable states from outside, the Gilbert algorithm explores the interior of the set, but by finding a witness, the set is also bounded from outside. Secondly, while the computational cost is intrinsic to recognition of entanglement, the hierarchy of PPTSE requires growing Hilbert spaces, which would eventually consume resources of any realistic computer. The Gilbert algorithm, on the other hand, works in the Hilbert space of our test state, and is limited only by computational time.

The algorithm can find many interesting applications. It can be combined with other algorithms, for example, for finding bound entangled states, or violation of Bell inequalities. This algorithm has the potential of greatly improving our understanding of the geometry of quantum states. 

We gratefully acknowledge O. G\"uhne for his help and support in writing this work. This work is a part of NCN grants No. 2014/14/E/ST2/00020 
and UMO-2017/26/E/ST2/01008.

\end{document}